\documentclass[%
 aip,
 amsmath,amssymb,
 reprint,%
]{revtex4-1}

\usepackage{graphicx}
\usepackage{dcolumn}
\usepackage{bm}
\usepackage[utf8]{inputenc}
\usepackage[T1]{fontenc}
\usepackage{mathptmx}
\usepackage{etoolbox}
\usepackage{color,soul}
\makeatletter
\def\@email#1#2{%
 \endgroup
 \patchcmd{\titleblock@produce}
  {\frontmatter@RRAPformat}
  {\frontmatter@RRAPformat{\produce@RRAP{*#1\href{mailto:#2}{#2}}}\frontmatter@RRAPformat}
  {}{}
}%
\makeatother
\begin{document}

\preprint{AIP/123-QED}

\title[Overcoming contrast reversals in focused probe ptychography of thick materials]{Overcoming contrast reversals in focused probe ptychography of thick materials: an optimal pipeline for efficiently determining local atomic structure in materials science}
\author{C. Gao}\author{C. Hofer}\author{D. Jannis}\author{A. B\'ech\'e}\author{J. Verbeeck}\author{T. J. Pennycook}
\affiliation{EMAT, University of Antwerp, 2020 Antwerp, Belgium }
\affiliation{NANOlab Center of Excellence, University of Antwerp, 2020 Antwerp, Belgium}

 \email{timothy.pennycook@uantwerpen.be}

\date{\today}

\begin{abstract}
Ptychography provides highly efficient imaging in scanning transmission electron microscopy (STEM), but questions have remained over its applicability to strongly scattering samples such as those most commonly seen in materials science. Although contrast reversals can appear in ptychographic phase images as the projected potentials of the sample increase, we show here how these can be easily overcome by a small amount of defocus. The amount of defocus is small enough that it can exist naturally when focusing using the annular dark field (ADF) signal, but can also be adjusted post acquisition. The ptychographic images of strongly scattering materials are clearer at finite doses than other STEM techniques, and can better reveal light atomic columns within heavy lattices. In addition data for ptychography can now be collected simultaneously with the fastest of ADF scans. This combination of sensitivity and interpretability  presents an ideal workflow for materials science. 
\end{abstract}

\maketitle

The ability of STEM to determine structure and composition has made it essential to materials science. Typically this has been via Z-contrast ADF imaging\cite{PENNYCOOK198958,PENNYCOOK199114,Nature_ORNL}, with simultaneous spectroscopies providing greater compositional sensitivity at the cost of far higher doses and drift\cite{EELS,PRL_Wu}. One particularly prominent deficiency of ADF imaging is the difficulty the modality faces in resolving light atoms near heavy elements. The precise locations of such light elements can dramatically alter the properties of materials containing them. Such is the case in many oxides where tiny changes in bond angles or lengths can completely change their magnetic or electronic properties\cite{light-atoms,Oxides,light-atoms1}.  Given sufficient stability of the stage and the sample, electron energy loss spectroscopy (EELS) can reveal the locations of such light elements\cite{PRL_Wu,EELS}. However, drift remains a challenge for precise measurements of atomic locations in EELS, and for many materials systems the damage from the doses required for spectroscopy preclude its use at atomic resolution. In many materials, the propensity to damage under the electron beam also precludes atomic resolution ADF imaging\cite{Beam_damage}. Alternative means of imaging with both greater sensitivity to light elements and higher overall dose efficiency have thus been sought to complement ADF imaging\cite{4D_STEM_beamsensitive,ADF_scan_strategies}. 

Annular bright field (ABF) became popular because it can often determine the locations of light elements in heavy lattices at higher precision and at lower doses than spectroscopic elemental mapping\cite{ABF2,ABF_H,ABF_findlay,ABF1}. Recently methods based on tracking the center of mass (CoM) of the electron scattering have begun to take over from ABF, as they are able to provide a higher signal to noise ratio. Integrated differential phase contrast (iDPC)\cite{idpc} and integrated CoM (iCoM)\cite{LAZIC2016265,iCOM_jordan} have shown particular promise in this regard. iDPC can be performed using a set of conventional detectors arranged into quadrants or segments\cite{idpc,idpc_1}, while iCoM is performed using four dimensional STEM (4D STEM) \cite{ophus_2019} data. Thus while iCoM utilizes a more accurate measure of the CoM, most cameras used for 4D STEM have made it significantly slower than iDPC. Now however developments in camera technology have made 4D STEM possible without any decrease in scan speed compared to even the most rapid ADF imaging\cite{Timepix3}. This technological advance also accelerates the acquisition speed for highly efficient ptychography\cite{Rodenburg1993304,SSB,Jiang_ptycho,ptycho3D}. Direct focused probe forms of ptychography are compatible with simultaneous ADF imaging, but their applicability to stronger objects such as thicker samples containing heavier elements have remained in question due to the use of the weak phase object (WPO) or multiplicative approximations in the theory underlying these ptychographic methods\cite{SSB2,WDD1,WDD2,SSB,SSB_dose_efficiency}.

Here we demonstrate the significant advantages of combining focused probe ptychography with conventional rapid scan ADF workflows for the thicker samples typical of materials science. Both the single side band (SSB) and Wigner distribution deconvolution (WDD) methods are capable of providing significantly clearer images of thick structures at lower doses than ABF and iCoM. This is true well beyond the approximations used in the theoretical description of the methods. Although complex thickness induced contrast reversals can appear in the SSB and WDD phase images, as has previously been shown\cite{WDD2}, we show how defocus adjustment can remove these effects and provide contrast reversal free images that can clearly show the positions of all the atomic columns in thick structures. The amount of defocus required can be within the range of defocus that is set naturally when optimizing ADF images manually in conventional imaging, but it can also be applied and optimized after acquiring the data due to the ability of ptychography to perform post collection aberration correction.

In SSB and WDD ptychography the phases of diffracted convergent beam electron diffraction (CBED) disks are solved via their mutual interference with the direct beam in probe position reciprocal space. Experiments proceed by recording the intensity distribution of the scattering as a function of probe position with a camera. The Fourier transform of this 4D STEM data with respect to probe position yields, in the multiplicative approximation\cite{WDD1},
\begin{equation}\label{ptycho_Eq}
G(\mathbf{K}_f,\mathbf{Q}_p)=A(\mathbf{K}_f)A^*(\mathbf{K}_f+\mathbf{Q}_p)\otimes_{\mathbf{K}_f}\Psi_s(\mathbf{K}_f)\Psi_s^*(\mathbf{K}_f-\mathbf{Q}_p),
\end{equation}
in which $\mathbf{K}_f$ is the scattering vector leading to a location on the camera, $\mathbf{Q}_p$ is spatial frequency, $A$ is an aperture function expressing both the effect of the aperture size and the aberrations that appear within it, and $\Psi_s$ represents the Fourier transform of the specimen transmission function. This represents the convolution between the region of mutual overlap of the shifted and unshifted aperture functions and the interference of the diffracted beams differing in frequency by $\mathbf{Q}_p$. In SSB ptychography the WPO approximation is invoked to further simplify this to \cite{Rodenburg1993304}
\begin{eqnarray}\label{ssb_Eq}
G(\mathbf{K}_f,\mathbf{Q}_p)&=&|A(\mathbf{K}_f)|^2\delta(\mathbf{Q}_p)\nonumber\\&&+A(\mathbf{K}_f)A^*(\mathbf{K}_f+\mathbf{Q}_p)\Psi_s^*(-\mathbf{Q}_p)\nonumber\\&
&+A^*(\mathbf{K}_f)A(\mathbf{K}_f-\mathbf{Q}_p)\Psi_s(\mathbf{Q}_p),
\end{eqnarray}
from which it can be understood how to obtain the phase and amplitude of each frequency of the specimen transmission function from the double disk overlap regions where the shifted and unshifted aperture functions coincide\cite{SSB}.  In WDD the WPO approximation is avoided and deconvolution is used to separate the aperture functions from the specimen transmission function\cite{WDD1}. However the principle remains essentially the same in that the phase and amplitude of each spatial frequency relative to the direct beam is determined from the double overlap regions. 

\begin{figure}
 \includegraphics[width=0.48\textwidth]{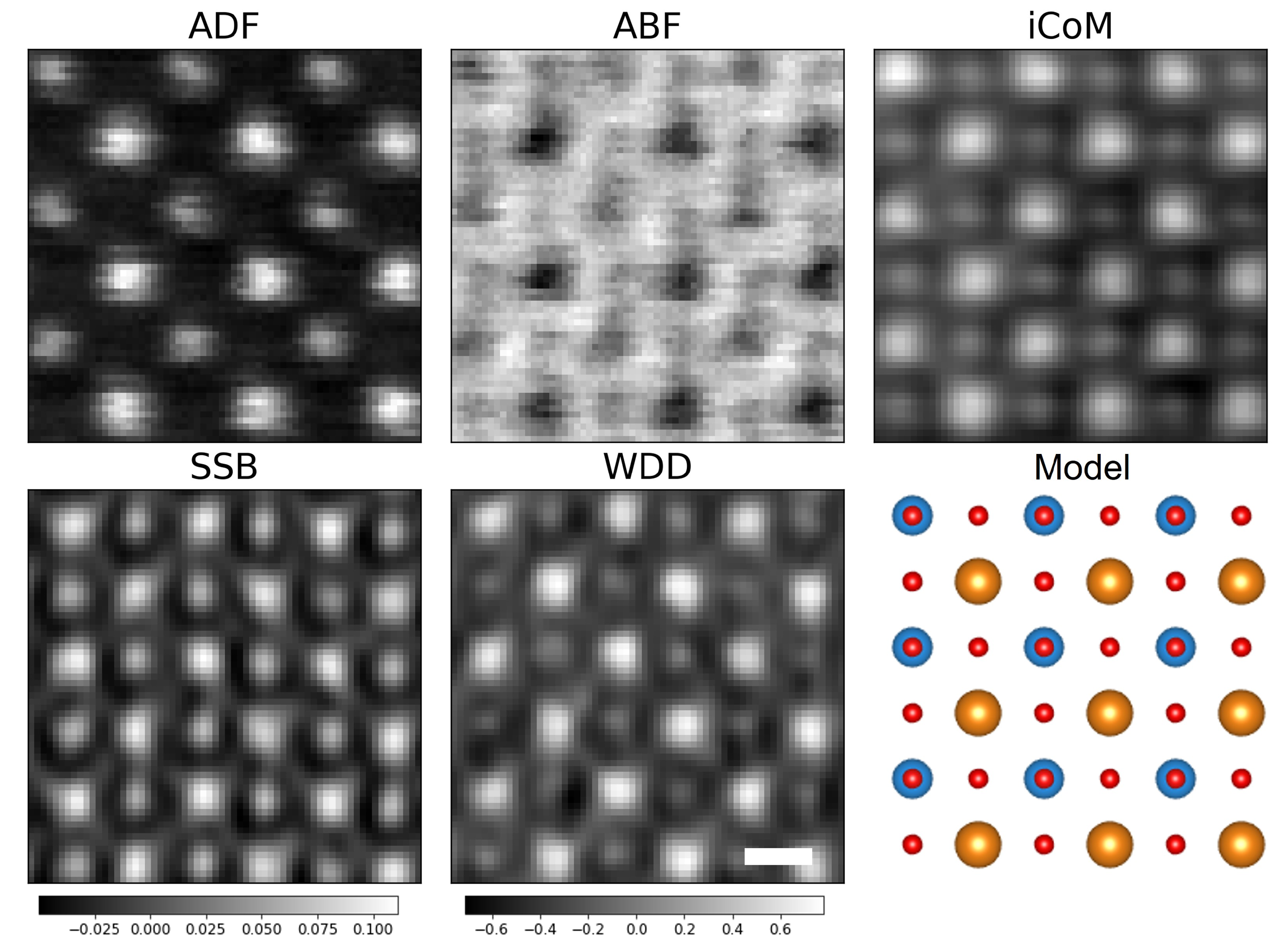}%
 \caption{\label{sto_merlin} ADF and Medipix3 based 4D STEM data from SrTiO$_3$. iCoM and ptychography show the O columns more clearly than the ABF. However this data shows some drift from the slow scan imposed by the camera. Sr, Ti and O columns are indicated by orange, blue and red respectively in the overlays. The scale bar shows 2 \AA~and the colorbars are in radians.}
 \end{figure}

Although the details of CBED patterns are dependent on thickness, both the WDD and SSB can in practice be seen to provide clear images of the structures of materials containing heavy elements that are likely tens of nanometers thick. An experimental example using a Medipix3 camera\cite{Merlin1} is shown with SrTiO\textsubscript{3} in Figure 1 comparing simultaneous ADF and 4D STEM based ABF, iCoM and the focused probe SSB and WDD ptychography signals. The thickness is most likely on the order of tens of nanometers from the position averaged CBED \cite{lebeau_position_2010} data (see supplementary information) which puts this sample well beyond both the WPO and multiplicative approximations. The conditions were optimized for the ADF image during the acquisition at 200 kV with a 20 mrad convergence angle. Distortions due to drift are apparent due to the very slow speed of the scan compared to conventional ADF imaging. The 2400 frames per second of the 6 bit mode of the Medipix3 camera used here is a typical speed for 4D STEM cameras over the past five years, and with a typical probe current of a few tens of picoamps results in a dose on the order of $10^6$ $e^-/$\AA$^2$. However it is clear that both SSB and WDD provide atomic resolution images that, along with the iCoM, are far clearer than that of the ABF, despite the strength of the scattering here. A degree of filtering is apparent, but most of the clarity is attributed to the dose efficiency of the iCoM and ptychographic methods. Note that the ADF remains useful for clearly distinguishing the Sr and Ti columns, but the O atoms are not visible without the additional signals. 

Frame based direct detection 4D STEM cameras such as the Medipix3 and its competitors are faster and more efficient than previous generation cameras, and are continuing to improve, but the susceptibility to instabilities such as 
drift and jitter resulting from their relatively low speeds remains a major problem for precision measurements. Furthermore the slow scan speeds they impose also make it more difficult to reach low doses. These two issues, instabilities and dose, have thus motivated continued use of methods based on conventional STEM detector setups such as ABF and iDPC which are in many ways inferior to 4D STEM methods but which have so far been more practical\cite{DPC,SSB,SSB2,LAZIC2016265}. 

Now we can remove the camera as a bottleneck to the speed at which 4D STEM can be performed. With event based detectors such as the Timepix3 \cite{Timepix3}, we can easily reach the equivalent of not just a few thousand frames per second, but millions of frames per second. We refer to the frames per second equivalent here in relation to the event driven camera only for the ease of comparison it can provide, as event driven operation does not utilize camera frames. Instead, each electron detected by the camera is labeled with a pixel location and time, and read out directly after the hit occurs rather than waiting for a full camera frame to be exposed and read out together. This provides a more efficient readout that enables us to easily perform 4D STEM at the single to few microsecond dwell times used in rapid scan ADF imaging. Faster scans are also possible, such as the 100 ns dwell time we demonstrated previously \cite{Timepix3}, and may become increasingly desirable with the enhanced dose efficiency available with CoM and especially ptychography. Indeed we can also perform the same type of drift corrected multiple scan based imaging as has become common in rapid scan ADF imaging, allowing us to use data up until the point damage sets in. Thus with event driven cameras 4D STEM can now be acquired as rapidly as any other STEM modality. 

\begin{figure}
 \includegraphics[width=0.48\textwidth]{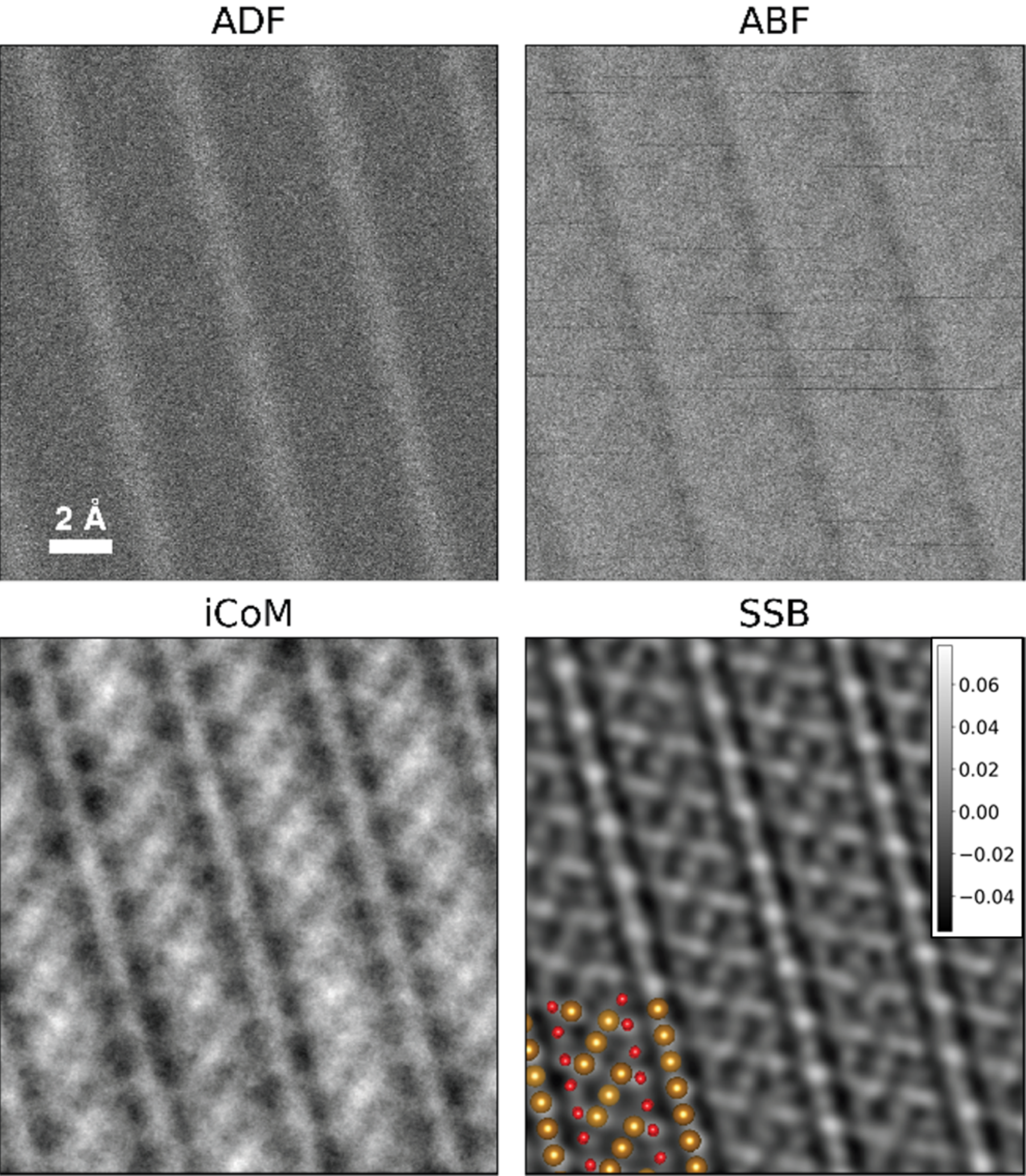}%
 \caption{\label{fe3o4} ADF and 4D STEM ABF, iCoM and SSB ptychography images of Fe\textsubscript{3}O\textsubscript{4} from the same 2 $\mu$s scan enabled by a Timepix3 camera. Despite the low dose, the SSB image clearly shows the oxygen columns. The crystallinity and absence of distortions allowed the full image to be divided into 4 parts and aligned, based on the SSB signal, and averaged to produce the higher signal-to-noise ratio images shown. O and Fe atoms are shown in red and brown respectively in the  model. The projection vector of the plane is (2,3,1) and the colorbar for the SSB is in radians.}  
 \end{figure}
 
We illustrate the step change in the quality of images microsecond dwell time 4D STEM provides in Figure 2 with data acquired on a Timepix3 complementing ADF data collected simultaneously from Fe\textsubscript{3}O\textsubscript{4}, again using 200kV and a 20 mrad convergence angle. The electron dose was 10000 $e^-/$\AA$^2$ and the dwell time was 2 microseconds. The ADF signal was used to optimize the imaging conditions, just as in a standard high end ADF STEM workflow. From both the Ronchigram and ADF imaging the sample is qualitatively of a thickness typical of 3D materials science samples for STEM. As seen in Figure~\ref{sto_merlin} the clarity of the iCoM and focused probe ptychography far exceed that of the ABF image. However unlike Figure~\ref{sto_merlin}, this scan is essentially free of drift and other instabilities and at approximately two orders of magnitude lower dose. Because of the relatively low dose employed, the ADF is extremely noisy, as is the ABF signal which provides little advantage over the ADF. With the lower dose used here, the strength of the focused probe ptychography over that of iCoM also becomes much more apparent. Essentially every single atomic column is visible in the focused probe ptychography, but they are not in the iCoM. 
Although the contrast transfer functions of SSB and WDD ptychography are single signed and provide easily interpretable images in which the contrast of the atoms is consistent against the background for thin weak specimens\cite{SSB2,Column}, the contrast becomes more complex when handling thicker samples. Complex changes in contrast can appear as the projected potential of the sample becomes stronger as evidenced previously in GaN\cite{WDD2}. Atomic columns can be seen to change from bright to dark as the sample thickens, generally beginning in the center of the atomic columns where the projected potential is strongest. Thus one can encounter images in which the center of atomic columns have become dark while further out they remain bright. Such complex contrast reversals can not only impede the ease of interpretability but also degrade the dose efficiency by reducing the available contrast in low dose images.
\begin{figure}
 \includegraphics[width=0.48\textwidth]{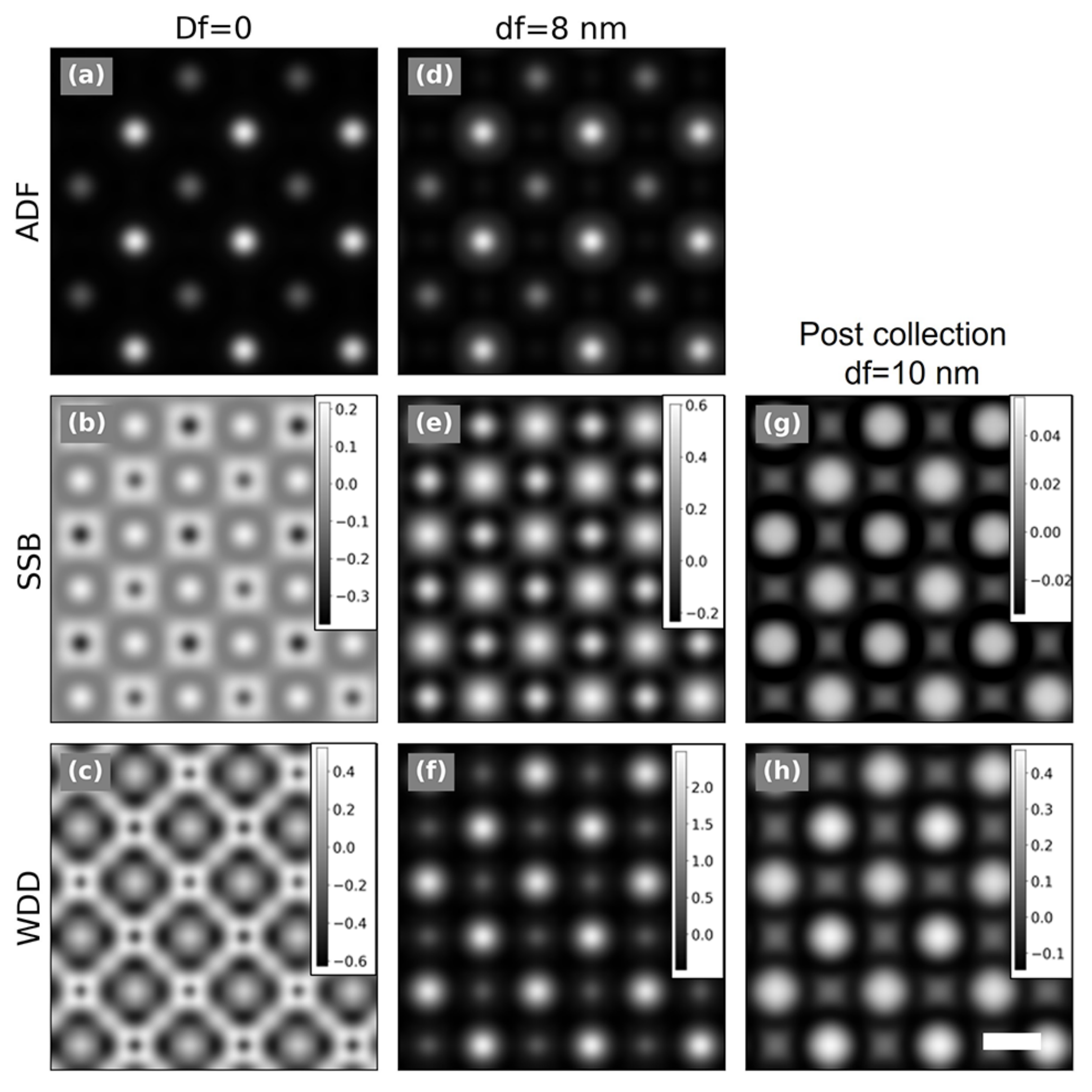}%
 \caption{\label{stosims} ADF, SSB and WDD images simulated for 16 nm thick STO. At zero defocus complex contrast reversals appear in the SSB and WDD images, but when the probe is focused 8 nm into the material the contrast reversals are removed. Importantly however the contrast reversals can also be removed by inserting defocus post collection, meaning the experiment can proceed with the focus optimized for the ADF. The scale bar indicates 2 \AA~and the colorbar is in radians.}
 \end{figure}
An example of such contrast reversals is shown with simulations in Figure \ref{stosims} (b) and (c). The 4D datasets are simulated using the multislice method with abTEM \cite{abTEM}. The voltage and convergence angle are the as same as the experimental results shown in Figure \ref{sto_merlin}. With the probe focused to the entrance surface of the 16 nm thick STO (zero defocus), both the SSB and WDD produce complex phase images in which the centers of both the Ti and Sr columns are dark but the light O columns remain bright. Surrounding the dark centers of the Ti and Sr columns are regions of higher phase values before these reduce to an intermediate phase value in the background. It is still possible to interpret which columns are where, particularly with the ADF signal present and indeed, the O columns already appear bright. Thus it is already feasible to locate the O atoms with this combination. However, these complex contrast reversals are clearly often undesirable, and by applying a small amount of defocus it is possible to remove them as illustrated in Figure 3 for SSB (e) and WDD (f) with a defocus of 8 nm, which has a minimal effect on the ADF image. With this defocus all the atomic columns appear bright on a dark background in the ptychographic images. Note that we define positive defocus as underfocus, moving the focal point into the sample from the entrance surface, and negative values as overfocus, moving the focal point before the sample. Focal series for thicknesses of 16, 20, 28 and 50 nm are shown in the supplementary materials. The optimal defocus appears to be close to the middle of the sample for the thinner examples, but for the 50 nm case defoci corresponding to a few nm before the sample or near the exit surface appear preferable. 

In Figure 3 we also show the effect of adding 10 nm of defocus post collection to the SSB (g) and WDD (h) for the 16 nm thick STO simulation. 10 nm was chosen as it optimizes the removal of the contrast reversals. This also works for considerably thicker materials as we show in the supplemental materials with a focal series for the 50 nm thick STO simulation. Using post collection SSB defocus adjustment we are able to remove the complex contrast reversals with the probe focused to the entrance surface during the simulated scan (with the CBED patterns simulated with defocus equal to zero). 
Although not identical to performing the defocus adjustment on the probe directly, this illustrates that post acquisition adjustment can also be used to remove the contrast reversals. The difference between applying the defocus during and after acquisition could be explained by channeling effects which are not accounted for with the post acquisition adjustment of the defocus. 

The availability of the simultaneous ADF signal is thus of two fold importance. Its atomic number contrast provides greater compositional discernment without resorting to the greater doses required for atomic resolution spectrum imaging, while providing a reference image which is not susceptible to contrast reversals. Atoms are always bright on a dark background in ADF images. This advantage is not available in defocused probe methods of ptychography because the ADF image is blurred beyond usability. 

Although elucidating the physics of the removal of contrast reversals via defocus in detail is reserved for future studies, we believe the effect can be understood as rolling the phase around across the field of view such that the centers of the atomic columns are maximum and the background at a minimum. This works in practice because the ptychographic phase images generally only use a small fraction of the $2\pi$ radian range of values available for phase value images, and thus the phase rolling does not bring other features into non intuitive ranges of contrast. 

In our experimental examples, the ptychographic data did not need additional defocus to produce clear images. This could be because they were obtained at a focus that produced clear ADF and ptychographic images. The defocus applied to provide contrast reversal free ptychographic images in Figure 3 only slightly reduced the quality of the ADF image. However when the defocus is not optimal for ptychography, it can be adjusted after taking the  data. The injection of defocus and post collection aberration correction has been demonstrated previously \cite{WDD1,Column} with focused probe ptychography, but here provides an unanticipated benefit. Post collection aberration correction can also be combined with such defocus optimization.

In comparison to direct ptychography, iterative ptychography is not guaranteed to converge to the correct result or indeed to converge at all, but they are guaranteed to take a longer time to produce a useful image. We timed serial SSB and extended ptychographic iterative engine (ePIE) \cite{maiden_improved_2009} calculations from the same focused probe dataset using an i7-10700K processor. The entire SSB calculation took 2.4 s regardless of dose. The ePIE calculation took 3.5 s (1 $\alpha$ limit) to 24.4 s (3 $\alpha$ limit) per iteration so that 10 ePIE iterations takes between 15 and 100 times as long as the SSB result. Convergence depends on the dose, and at $1.2\times 10^5e^-/$\AA$^2$ ePIE did not converge whereas the SSB produces clear images. Parallelism can speed up both algorithms, but this shows the far higher computational costs of iterative vs direct methods. See the supplementary information for further details. Live iCoM\cite{Live_iCoM} and SSB\cite{Live_SSB} have already been reported, but live imaging with iterative methods would be at least far more difficult. Detailed information about simulation methods and the related parameters can be found in the supplementary information.

In conclusion, we have demonstrated the utility of performing focused probe ptychography within existing rapid scan ADF workflows for general materials science samples. Simultaneous ADF and ptychography provide a robust and interpretable method of efficiently imaging all the atomic columns in such samples. In particular, when combined with the ability to apply corrective defocus, the method works well beyond the approximations which have been used to motivate the SSB and WDD methods. The clarity of the ptychographic images, provides the best possible combination of the general efficiency of structural imaging, visibility of light elements and interpretability when combined with simultaneous Z-contrast ADF imaging. With a detector such as the Timepix3, drift is no longer a significant issue with 4D STEM and there is indeed very little reason not to perform simultaneous ADF and 4D STEM imaging. We therefore anticipate the widespread adoption of this imaging pipeline in the future. 

\section*{Acknowledgements}
We acknowledge funding from the European Research Council (ERC) under the European Union’s Horizon 2020 research and innovation programme (C.G., C.H. and T.J.P.) via 802123-HDEM.
This project has received funding from the European Union’s Horizon 2020 Research Infrastructure - Integrating Activities for Advanced Communities under grant agreement No 823717 – ESTEEM3. J.V. and D.J. acknowledge funding from FWO project, Belgium G042920N "Coincident event detection for advanced spectroscopy in transmission electron microscopy" and J.V. and A. B. acknowledge funding from G042820N "Exploring adaptive optics in transmission electron microscopy". We acknowledge funding under the European Union’s Horizon 2020 research and innovation programme (J.V. and D.J) under grant agreement No 101017720 "eBEAM: Electron beams enhancing analytical microscopy".
\bibliography{aipsamp}

\end{document}


\title[]{Supplementary Information:\\ Overcoming contrast reversals in focused probe ptychography of thick materials: an optimal pipeline for efficiently determining local atomic structure in materials science}
\author{C. Gao}\author{C. Hofer}\author{D. Jannis}\author{J. Verbeeck}\author{T. J. Pennycook}
\affiliation{EMAT, University of Antwerp, 2020 Antwerp, Belgium }
\affiliation{NANOlab Center of Excellence, University of Antwerp, 2020 Antwerp, Belgium}

 \email{timothy.pennycook@uantwerpen.be}
\date{\today}
\maketitle
\subsection*{Additional simulation and reconstruction details}
\noindent The abTEM simulations used the Kirkland potential parameterization\cite{kirkland-parameters}. \\

\noindent The convergence angle, probe position spacing, and rotation angles used in the ptychographic reconstructions matched the data input to them. No additional parameters are needed for the SSB algorithm. For WDD we additionally used an epsilon value of 0.01 and for ePIE values of alpha = 0.1 and beta = 1 were used. \\

\noindent The SSB and WDD reconstructions were peformed using the pyPtychoSTEM code which we provide on gitlab at \url{https://gitlab.com/pyptychostem/pyptychostem}. \\

\noindent For ePIE we used the code available from Cornell at \url{https://github.com/paradimdata/Cornell_EM_SummerSchool_2021/tree/main/Tutorial%206%20-%20Ptychography}.

\begin{figure}
 \includegraphics[width=0.78\textwidth]{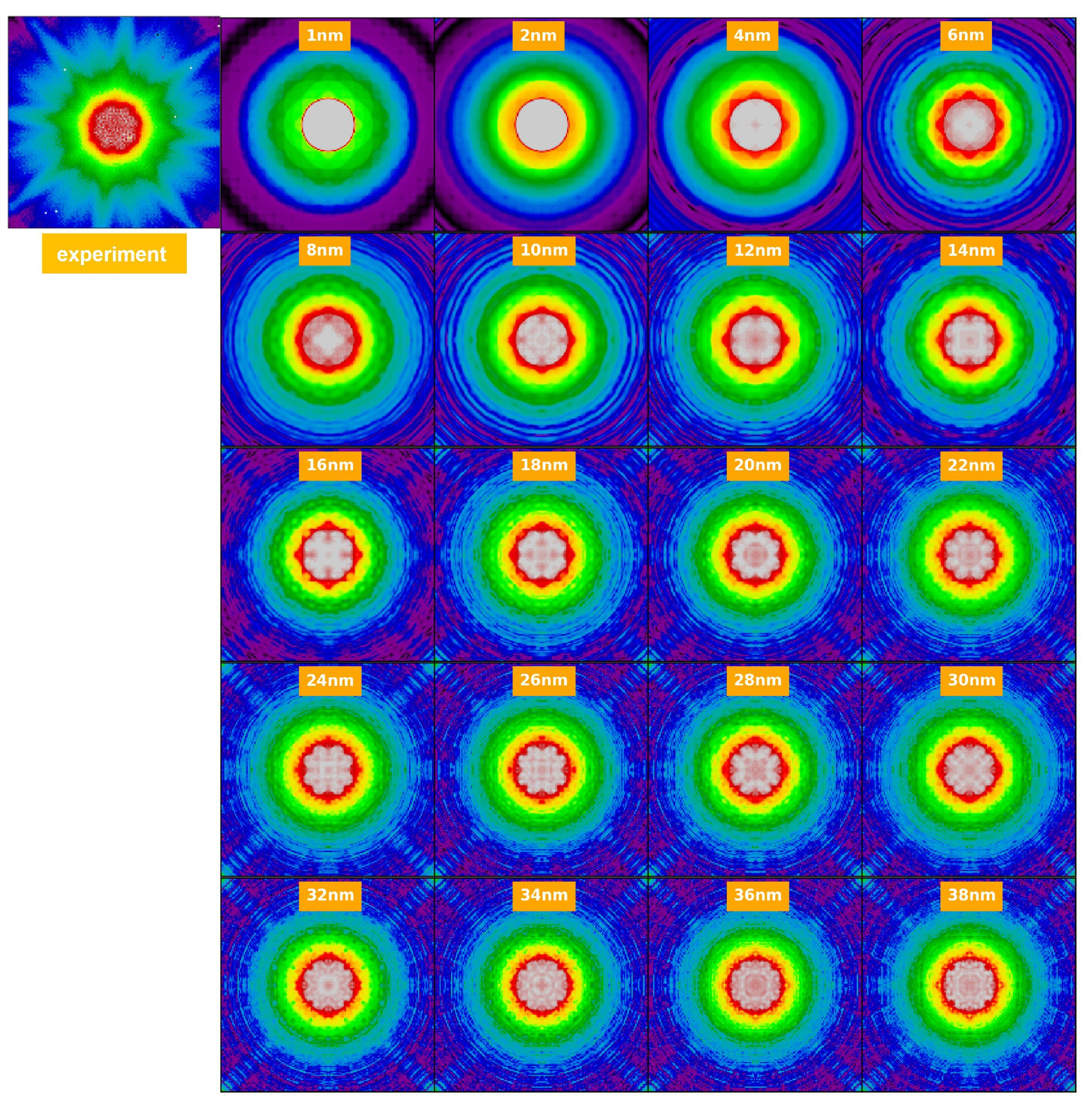}%
 \caption{\label{PACBED} Comparison of the position averaged convergent beam electron diffraction (PACBED) \cite{lebeau_position_2010}  from the experimental data used for Figure 1 of the main text with PACBED simulated for different STO sample thicknesses using the same 200 kV accelerating voltage and 20 mrad convergence angle used in the experiment. While no single thickness provides a perfect match, it is clear the experimental PACBED is consistent with the sample tens of nanometers thick. Looking at the symmetries present, we estimate the experiment was performed in a region that is a least approximately 20 nm thick. The scale bar is 2 \AA ngstroms. }
 \end{figure}

\begin{figure}
\includegraphics[width=0.78\textwidth]{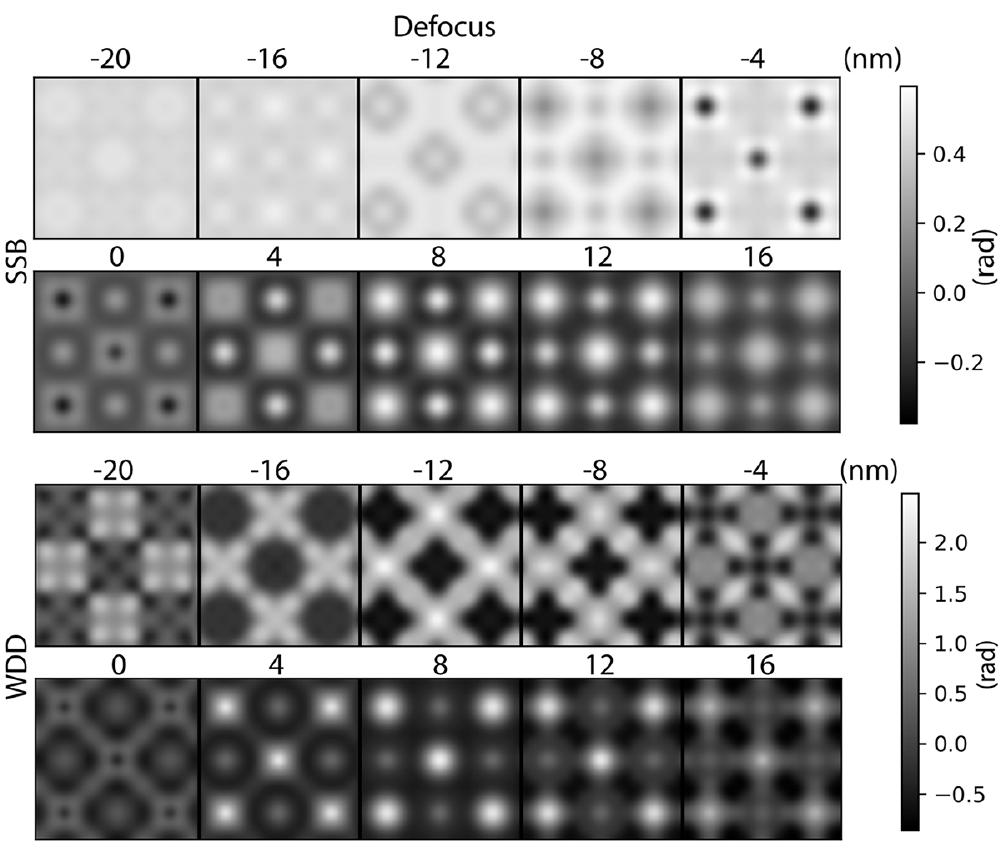}
\caption{\label{16 nm STO} Simulated SSB and WDD reconstructed phase results for 16 nm thick STO with varying probe defocus. }
    \label{fig:my_label}
\end{figure}

\begin{figure}
\includegraphics[width=0.78\textwidth]{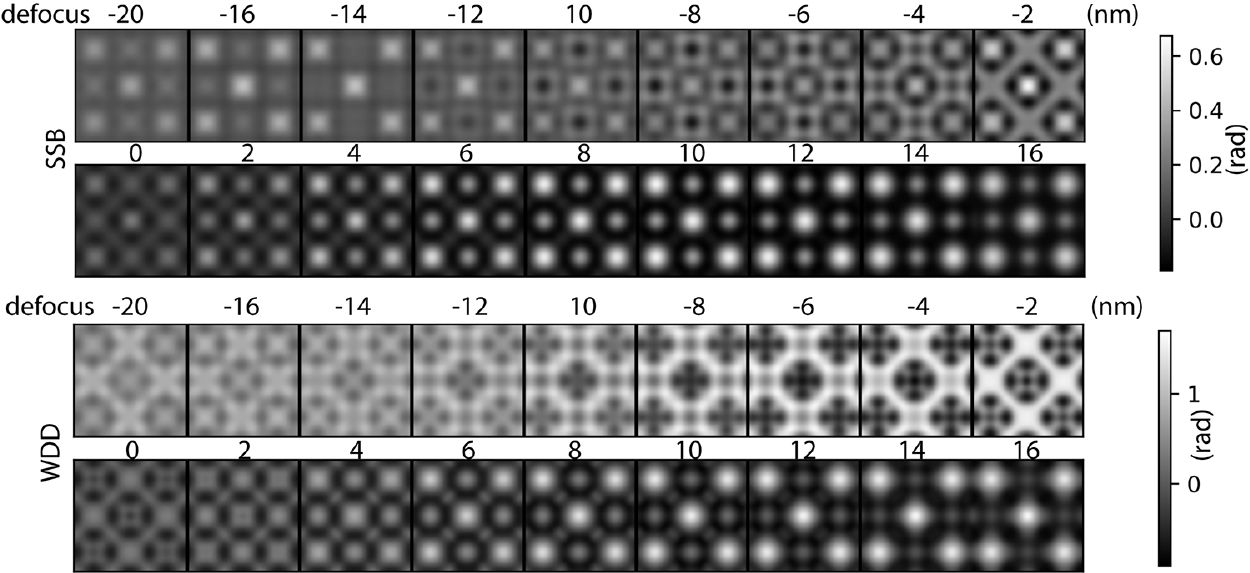}
\caption{\label{20 nm STO} Simulated SSB and WDD reconstructed phase results for 20 nm thick STO with varying probe defocus.  }
    \label{fig:my_label}
\end{figure}

\begin{figure}
\includegraphics[width=0.78\textwidth]{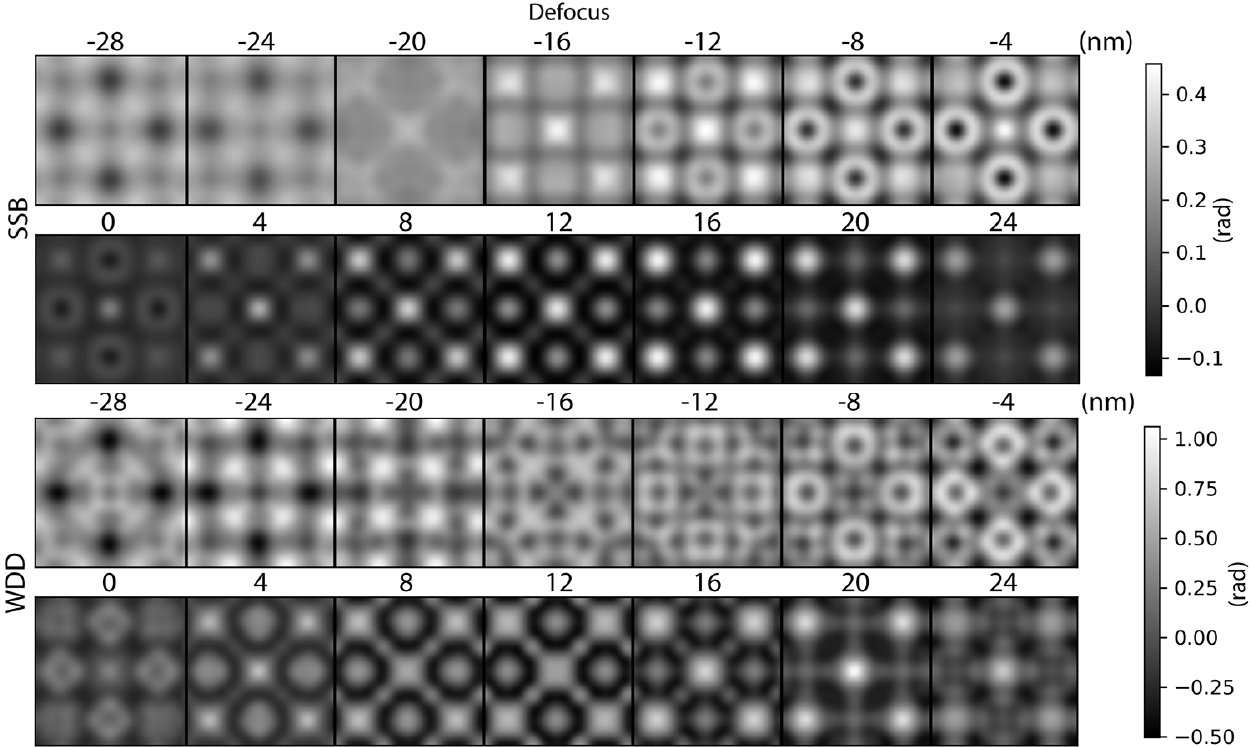}
\caption{\label{28 nm STO} Simulated SSB and WDD reconstructed phase results for 28 nm thick STO with varying probe defocus.  }
    \label{fig:my_label}
\end{figure}

\begin{figure}
\includegraphics[width=0.78\textwidth]{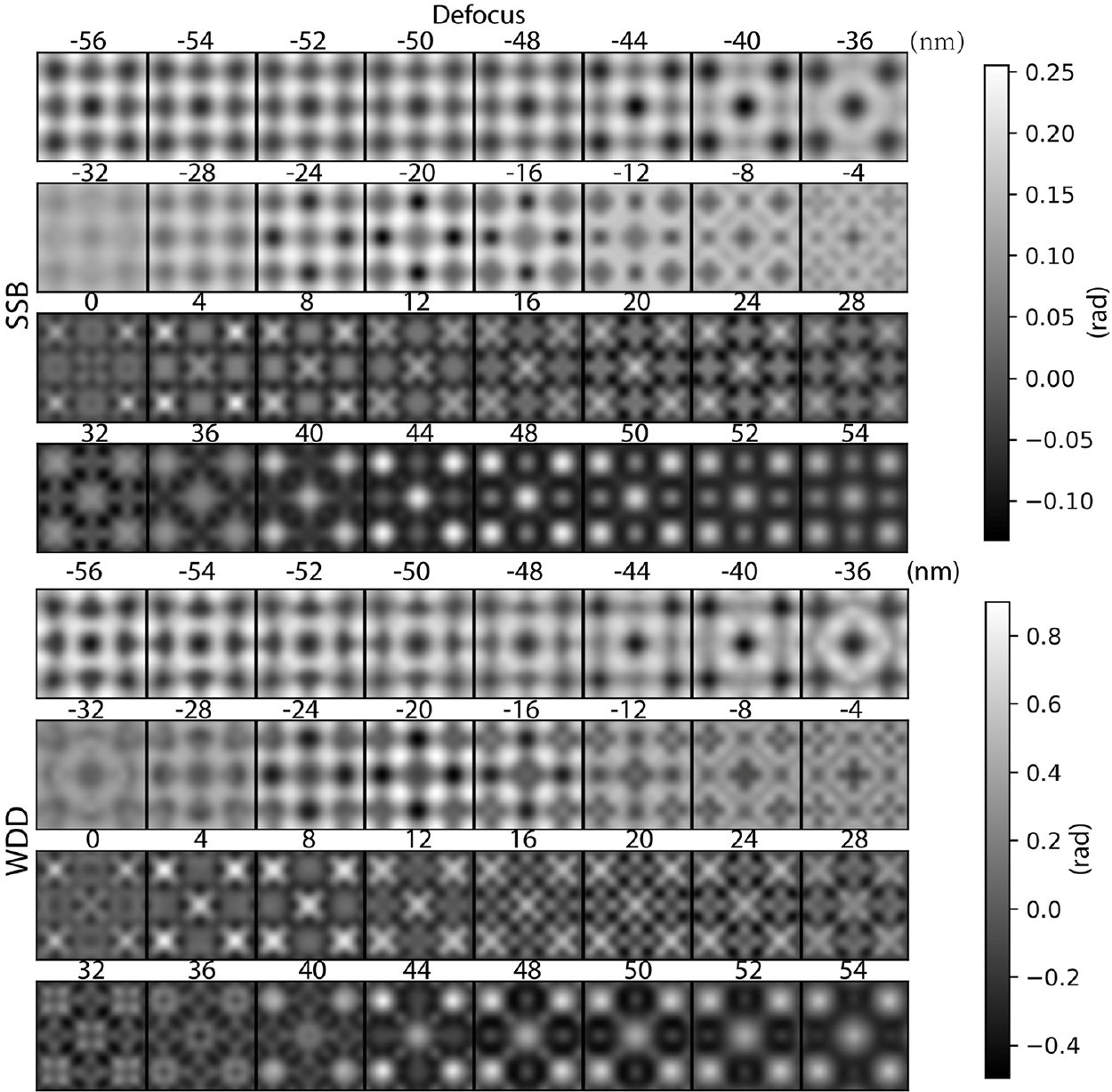}
\caption{\label{50 nm STO} Simulated SSB and WDD reconstructed phase results for 50 nm thick STO with varying probe defocus. }
    \label{fig:my_label}
\end{figure}

\begin{figure}
\includegraphics[width=0.78\textwidth]{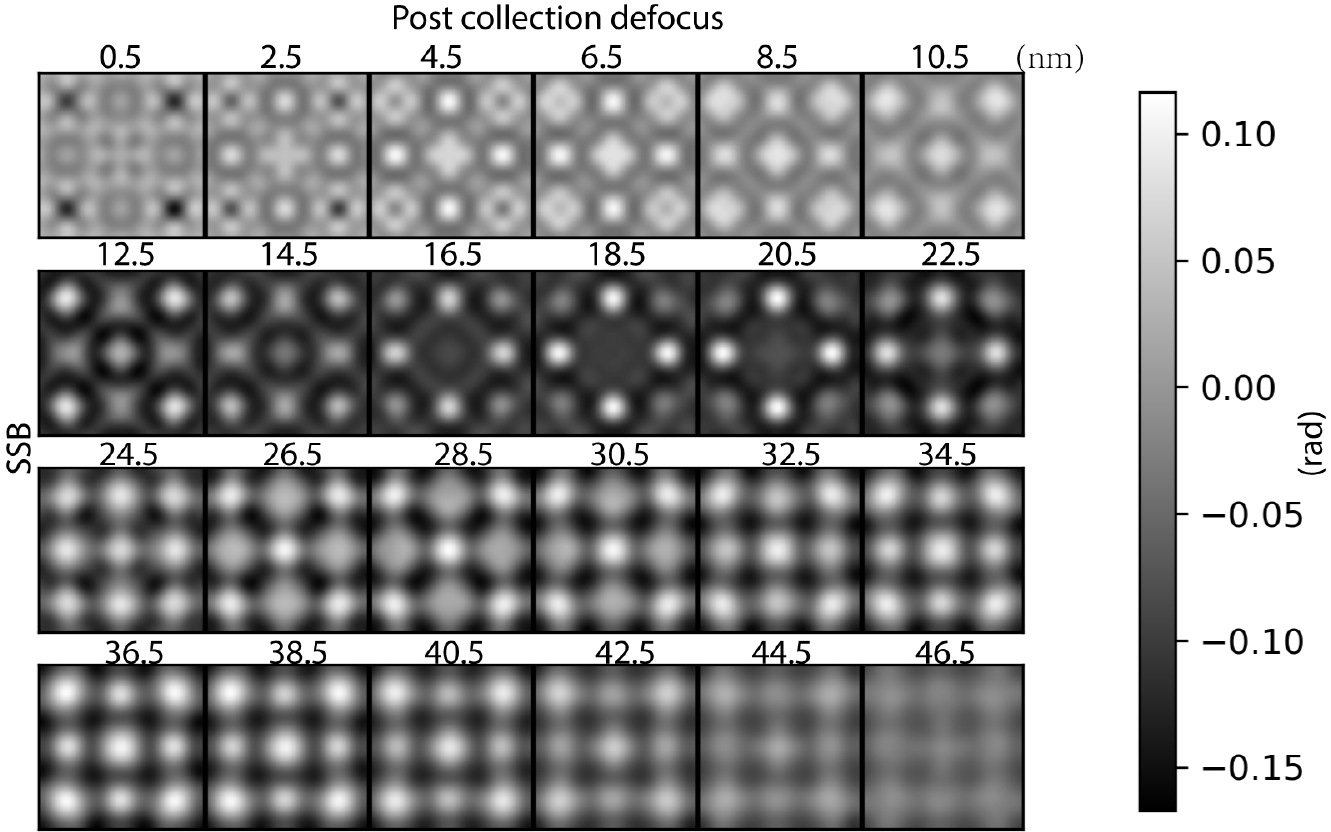}
\caption{\label{post50nmSTO} Simulated post collection SSB focal series for 50 nm thick STO, demonstrating that the contrast reversals can also be removed for this thickness when the probe is focused to the entrance surface during the scan with post collection defocus adjustment. }
    \label{fig:my_label}
\end{figure}

\begin{figure*}
 \includegraphics[width=0.78\textwidth]{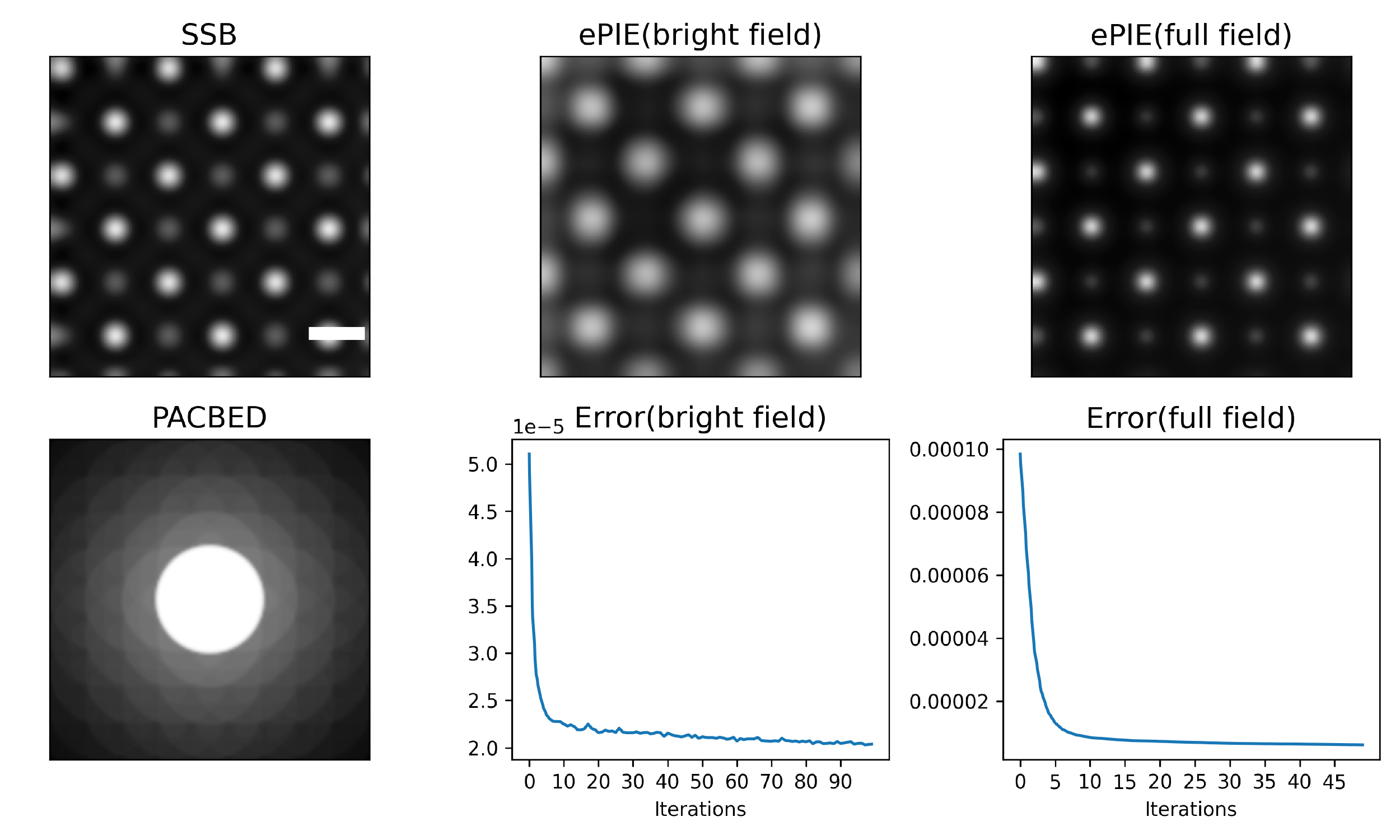}%
 \caption{\label{epievsssb} ePIE and SSB results computed using a single core of an i7-10700K operating at 3.8 GHz using data simulated for a single unit cell of STO at 200 KV, with a 20 mrad convergence angle, zero defocus and a 59 by 59 probe position grid. The 47 by 47 pixels of the bright field disk are used for bright field ePIE, while the 137 by 137 pixels of the full field CBED patterns are used for the full field ePIE. The SSB image was computed in a total time of 2.4 s. 100 iterations were used in computing the bright field ePIE result taking a total time of 347 s, for an average of 3.47 seconds per iteration. The full field ePIE result and the error per iteration are also provided. The PACBED of the STO dataset is also shown with a log scale.}

 \end{figure*}
\begin{figure}
\includegraphics[width=0.78\textwidth]{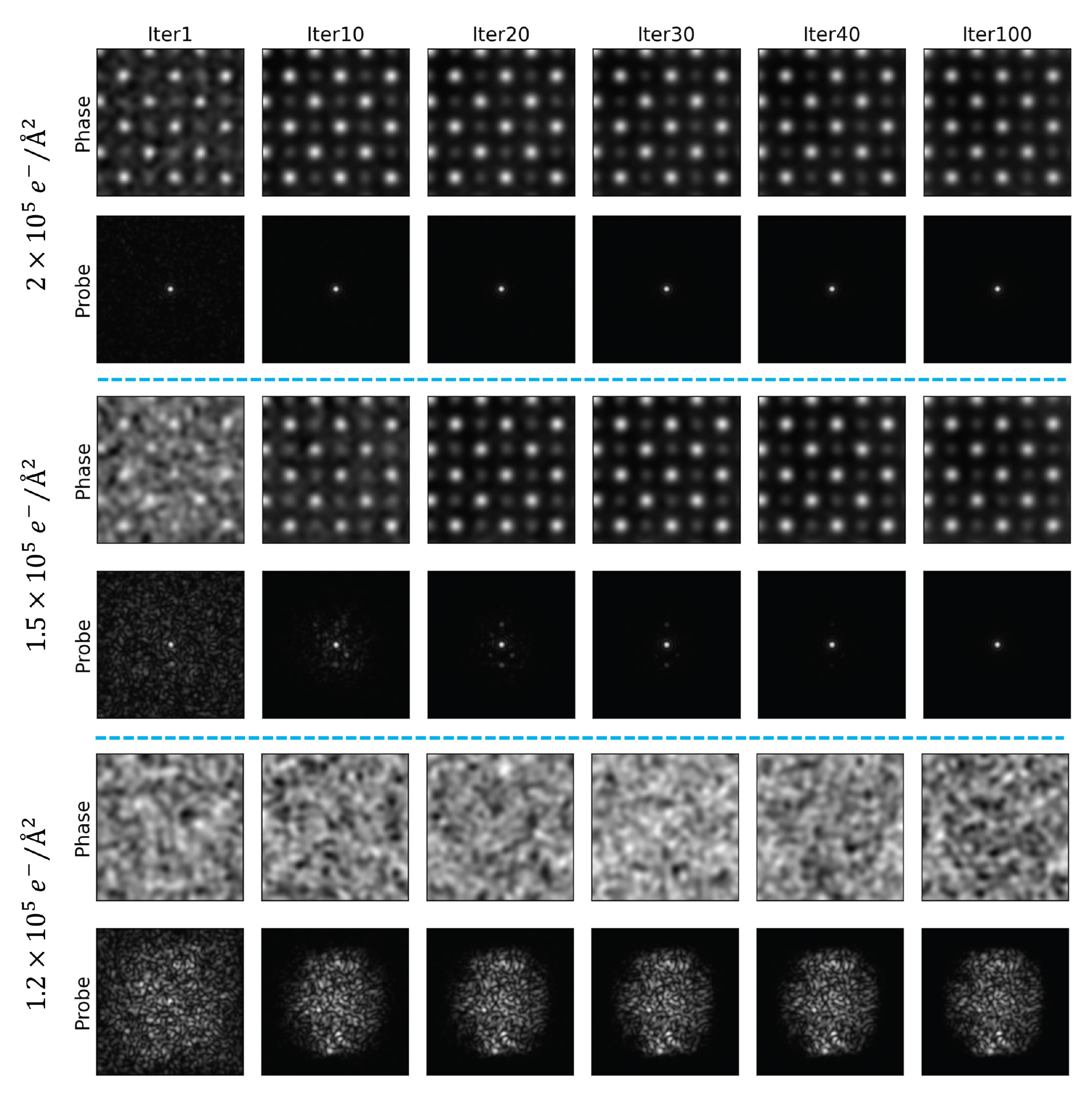}
\caption{\label{ePIE} Full field single unit cell thick STO ePIE reconstruction as a function of iteration number at different doses. At $2\times 10^5e^-/$\AA$^2$, after 10 to 20 iterations very little apparent change occurs in the phase and probe. At $1.5\times 10^5e^-/$\AA$^2$ some artefacts are visible at 20 iterations, but disappear between iterations 30 to 40. At  $1.2\times 10^5e^-/$\AA$^2$ the algorithm does not converge even after 100 iterations, reflecting the fact that the convergence of ePIE depends critically on the amount of signal in the CBED patterns. We note that using a significantly defocused probe increases the signal per CBED pattern, but a strongly defocused probe precludes useful simultaneous Z-contrast ADF imaging.}
    \label{fig:my_label}
\end{figure}

\begin{figure}
\includegraphics[width=0.78\textwidth]{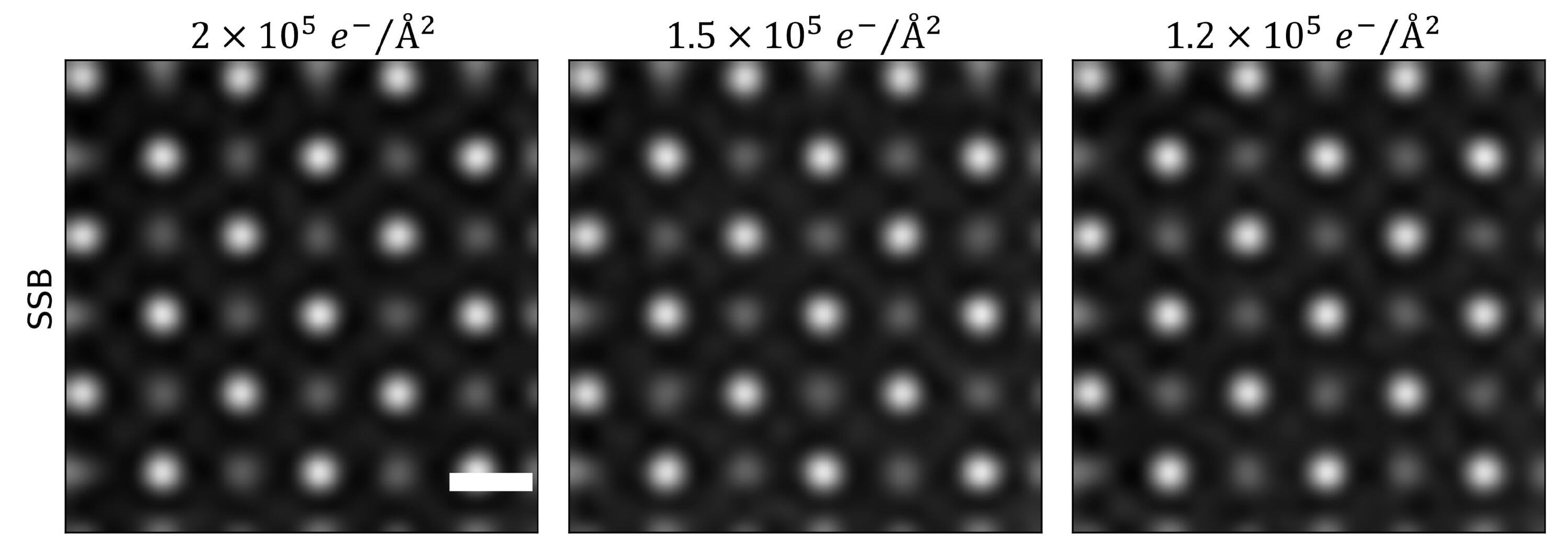}
\caption{\label{SSB} SSB reconstructions at the same doses as the ePIE results shown in Figure S8. As a direct non-iterative method, SSB ptychography does not depend on the dose to converge. Of course, the clarity of the resulting SSB images do depend on the dose, but it is a consistent and immediate result, and the above doses pose no problem at all for the SSB method. Scale bar is 2 \AA ngstroms. }
    \label{fig:my_label}
\end{figure}


\bibliography{aipsamp}